\documentclass[useAMS,usenatbib]{mn2e}
\usepackage{graphicx}
\usepackage{amssymb}
\usepackage{hyperref}

\def\be{\begin{equation}} 
\def\ee{\end{equation}}

\def\msun{{\Msun}}

\def\HI{\hbox{H~$\scriptstyle\rm I\ $}} 
\def\HII{\hbox{H~$\scriptstyle\rm II\ $}}

\def\gsim{\lower.5ex\hbox{\gtsima}} 
\def\lsim{\lower.5ex\hbox{\ltsima}} \def\gtsima{$\; \buildrel > \over 
\sim \;$} \def\ltsima{$\; \buildrel < \over \sim \;$} \def\prosima{$\; 
\buildrel \propto \over \sim \;$} \def\gsim{\lower.5ex\hbox{\gtsima}} 
\def\lsim{\lower.5ex\hbox{\ltsima}} 
\def\simgt{\lower.5ex\hbox{\gtsima}} 
\def\simlt{\lower.5ex\hbox{\ltsima}} 
\def\simpr{\lower.5ex\hbox{\prosima}} \def\la{\lsim} \def\ga{\gsim} 
  
 \def\gtsima{$\; \buildrel > \over \sim \;$} 
\def\ltsima{$\; \buildrel < \over \sim \;$} 
\def\gsim{\lower.5ex\hbox{\gtsima}} 
\def\lsim{\lower.5ex\hbox{\ltsima}} 
\def\simgt{\lower.5ex\hbox{\gtsima}} 
\def\simlt{\lower.5ex\hbox{\ltsima}} 
\def\simpr{\lower.5ex\hbox{\prosima}} 
\def\la{\lsim} 
\def\ga{\gsim}

\def\Lya{Ly$\alpha$~}

\def\msun{\,{\rm \Msun}}

\def\E3{{\cal E}_{\rm g}^{III}}

\def\Msun{\rm M_\odot}

\def\fesc{$f_{\rm esc}$\,}
\def\lya{$L_\alpha$\,}
\def\highz{high-$z$\,}
\def\avchi{$\langle \chi_{HI} \rangle$}
 
\title[LAE clustering \& lifetime]{Clustering and lifetime of Lyman Alpha Emitters in the Epoch of Reionization}
\author[Hutter et al.]{Anne Hutter$^{1}$\thanks{E-mail: ahutter@aip.de}, Pratika Dayal$^{2}$, Volker M\"uller$^{1}$ \\ 
$^{{1}}$ Leibniz-Institut f\"ur Astrophysik, An der Sternwarte 16, 14482 Potsdam, Germany\\
$^{2}$ Institute for Computational Cosmology, Department of Physics, University of Durham, South Road, Durham DH1 3LE, UK}

\begin{document} 
 
\date{} 
 
 
\maketitle 
 
\label{firstpage} 
\begin{abstract} 
We calculate Lyman Alpha Emitter (LAE) angular correlation functions (ACFs) at $z \simeq 6.6$ and the fraction of lifetime (for the $100$~Myrs preceding $z\simeq6.6$)  galaxies spend as Lyman Break Galaxies (LBGs) or as LBGs with Lyman Alpha (Ly$\alpha$) emission using a model that combines SPH cosmological simulations (GADGET-2), dust attenuation and a radiative transfer code (pCRASH). The ACFs are a powerful tool that significantly narrows the 3D parameter space allowed by LAE \Lya and UV luminosity functions (LFs) alone. With this work, we simultaneously constrain the escape fraction of ionizing photons $f_{esc}=0.05-0.5$, the mean fraction of neutral hydrogen in the intergalactic medium (IGM) \avchi$\lsim 0.01$ and the dust-dependent ratio of the escape fractions of \Lya and UV continuum photons $f_{\alpha}/f_c=0.6-1.2$. Our results show that reionization has the largest impact on the amplitude of the ACFs, and its imprints are clearly distinguishable from those of $f_{esc}$ and $f_\alpha/f_c$. We also show that galaxies with a critical stellar mass of $M_* = 10^{8.5} (10^{9.5})\Msun$ produce enough luminosity to stay visible as LBGs (LAEs). Finally, the fraction of time during the past $100$~Myrs prior to $z=6.6$  a galaxy spends as a LBG or as a LBG with \Lya emission increases with the UV magnitude (and the stellar mass $M_*$): considering observed (dust and IGM attenuated) luminosities, the fraction of time a galaxy spends as a LBG (LAE) increases from 65\% to 100\% ($\simeq 0-100$\%) as $M_{UV}$ decreases from $M_{UV} = -18.0$ to $-23.5$ ($M_*$ increases from $10^8-10^{10.5}\Msun$). Thus in our model the brightest (most massive) LBGs most often show \Lya emission.
\end{abstract}


\section{Introduction}
The Epoch of Reionization (EoR) marks a major phase change in the ionization state of the Universe. While the intergalactic medium (IGM) is predominantly composed of neutral Hydrogen (\HI) at the beginning of this epoch, it is completely ionized by the end, as a result of \HI ionizing photons produced by both stars and quasars. However, the progress of reionization has been hard to pin down since it depends on a number of parameters including the initial mass function (IMF) of reionization sources, their star formation rates (SFR), their stellar metallicity and age, the escape fraction of \HI ionizing photons produced by each source and the clumping of the intergalactic medium (IGM), to name a few. A further complication is introduced by supernova feedback and (to a lesser extent) the ultraviolet background (UVB) built up during reionization in suppressing the gas content (and hence star formation) in low-mass galaxies which are the main sources of \HI ionizing photons \citep[see e.g.][and references therein]{barkana-loeb2001, ciardi-ferrara2005, maio2011, sobacchi2013, wyithe2013, dayal2015}.

Lyman Alpha (Ly$\alpha$) photons are a powerful tool in understanding the ionization state of the IGM given their high optical depth ($\tau$) to \HI \citep[e.g.][]{madau-rees2000}
\begin{equation}
\tau = 1.5 \times 10^5 h^{-1} \Omega_m^{-1} \frac{\Omega_b h^2} {0.019} \bigg(\frac{1+z}{8}\bigg)^{3/2} \left( 1+\delta_H \right)\ \chi_{HI},
\end{equation} 
where $h$ is the Hubble parameter, $\Omega_b$ and $\Omega_m$ represent the cosmic baryon and matter density, respectively, $(1+\delta_H)$ is the hydrogen over-density and $\chi_{HI}$ is the fraction of neutral hydrogen. As seen from this equation, even a neutral hydrogen fraction as low as $10^{-5}$ can lead to a significant attenuation of \Lya photons at \highz, making them extremely sensitive probes of \HI in the IGM. As a result, a class of \highz galaxies called Lyman Alpha Emitters (LAEs), detected by means of their \Lya line (at 1216\,\AA\, in the rest frame), have become popular probes of reionization, with statistically significant samples available in the reionization epoch, at $z \simeq 5.7$ and $6.6$ \citep{malhotra2005, taniguchi2005, shimasaku2006, hu2010, kashikawa2006, ouchi2010, kashikawa2011}. Indeed, a number of theoretical papers have used semi-analytic \citep[e.g.][]{dijkstra2007,dayal2008,samui2009} or numerical \citep[e.g.][]{mcquinn2007,iliev2008,dayal2011,forero2011,duval2014,hutter2014} models to reproduce the observed number counts of LAEs (the \Lya luminosity function; LF) at various redshifts in the reionization epoch. However interpreting a change in the \Lya LF is rendered challenging by the fact that the observed \Lya luminosity depends on: (a) the fraction of \HI ionizing photons (1-\fesc) produced by a galaxy that are able to ionize the interstellar \HI resulting in the \Lya recombination line, with the rest (\fesc) escaping to ionize the IGM, (b) the fraction of the intrinsic \Lya photons that can emerge out of the galactic environment unattenuated by dust ($f_\alpha$) and \HI and (c) the fraction of these emergent \Lya photons that are transmitted ($T_\alpha$) through the IGM (depending on \avchi) and reach the observer. 

Given these uncertainties, we require an alternative measurement to constrain the ionization state using LAEs. One such strong measure is provided by the two-point angular correlation function (ACF) of LAEs that describes their spatial clustering. Indeed, \citet{mcquinn2007} have shown that the spatial clustering can hardly be attributed to anything other than the large scale ($\sim$ Mpc scale) ionization regions created during reionization. This result is similar to that obtained by \citet{jensen2014} who find that upcoming large-field LAE surveys should be able to detect the clustering boost for sufficiently high global IGM neutral fractions ($20$\% at $z=6.5$), although \citet{jensen2013} point out that a sample of several thousand objects is needed to obtain a significant clustering signal.
Furthermore, \citet{zheng2010} have shown that \Lya radiative transfer modifies the ratio of observed and intrinsic \Lya luminosities depending on the density and velocity structure of the environment, i.e. LOS (transverse) density fluctuations are suppressed (enhanced), leading to a change in the amplitude of the two-point correlation function compared to the case without the environmental selection effect. However, while \citet{behrens2013} have confirmed a correlation between the observed \Lya luminosity and the underlying density and velocity field, they do not find a significant deformation of the two-point correlation function by post-processing hydrodynamical simulations with a \Lya radiative transfer code.  

A second probe is presented by the fraction of Lyman Break Galaxies (LBGs) that show \Lya emission: given that the physical properties of LBGs do not evolve in the 150~Myrs between $z \simeq 6$ and 7, a sudden change in the fraction of LBGs showing \Lya emission could be attributed to reionization \citep[e.g.][]{stark2010, stark2011, fontana2010, pentericci2011, caruana2014, faisst2014, schenker2014, tilvi2014}. However, this interpretation is complicated by caveats including the redshift dependence of the relative effects of dust on \Lya and ultraviolet (UV) continuum photons \citep{dayal2012} and the fact that simple cuts in EW and UV luminosity may lead to uncertainties in the LAE number densities \citep{dijkstra2012}.

Coupling a cosmological SPH simulation snapshot at $z \simeq 6.6$ with a radiative transfer (RT) code (pCRASH) and a dust model, \citet{hutter2014} have shown that the effects of \fesc, $f_\alpha$ and $T_\alpha$ are degenerate on the LAE visibility: reproducing the observed \Lya LFs can not differentiate between a Universe which is either completely ionized or half neutral (\avchi$\simeq 0.5-10^{-4}$), or has an \fesc ranging between 5-50\%, or has dust that it either clumped or homogeneously distributed in the interstellar medium (ISM) with $f_\alpha/f_c=0.6-1.8$; here $f_c$ represents the fraction of UV photons that emerge out of the ISM unattenuated by dust. In this work, we extend our calculations to use the two-point LAE ACF to narrow down the 3-dimensional parameter space of \avchi, \fesc and $f_\alpha/f_c$. Further, we study the time evolution of both the \Lya and UV luminosities to show the {\it fraction} of time during the past $100$~Myrs prior to $z=6.6$ that a galaxy is visible as a LBG or as a LBG with Ly$\alpha$ emission, both considering the {\it intrinsic} and {\it observed} (dust and IGM-attenuated) luminosities. In principle, the ratio of these fractions could be measured by relating the number of observed LBGs with Ly$\alpha$ emission to the number of observed LBGs at $z=6.6-7.3$.

The cosmological model corresponds to the $\Lambda$CDM Universe with dark matter (DM), dark energy and baryonic density parameter values of ($\Omega_{\Lambda}$, $\Omega_m$, $\Omega_b$) = (0.73, 0.27, 0.047), a Hubble constant $ H_0= 100h = 70 {\rm km s^{-1} Mpc^{-1}}$, and a normalisation $\sigma_8=0.82$, consistent with the results from WMAP5 \citep{komatsu2009}. 

\section{The model}
\label{sec_model}
In this section we briefly describe our physical model for high-redshift LAEs that couples cosmological SPH simulations run using GADGET-2 with a RT code (pCRASH, \citet{partl2011}) and a dust model, and interested readers are referred to \citet{hutter2014} for a detailed description. 

\begin{figure*}
  \center{\includegraphics[width=1.03\textwidth]{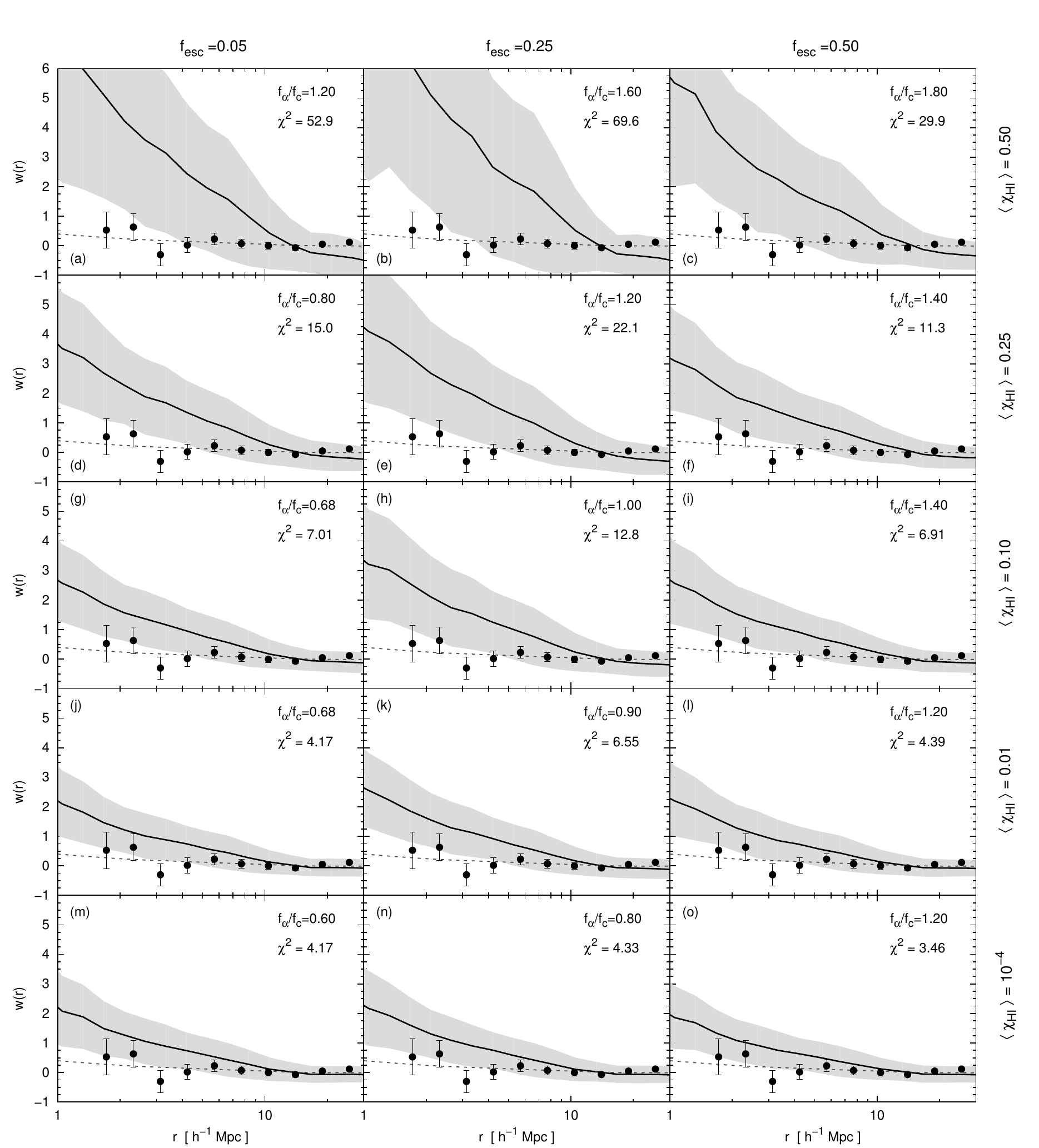}}
  \caption{Angular correlation function for simulated LAEs. The mean ACF is calculated from 36 mock catalogues (12 along each of $x,y,z$ directions) assuming volumes with a depth of $30 h^{-1}$ Mpc and a FoV of $\sim3\times 10^3 h^{-2}$ Mpc$^2$. In each panel, the solid line shows the ACF of the best fit $(f_{\alpha}/f_c$, \fesc, $\langle \chi_{HI} \rangle)$ combinations for which the simulated LAE \Lya LFs are within the $1\sigma$ limit of the observations by \citet{kashikawa2011} with shaded regions showing the variance across the mock catalogues. In each panel, the grey dashed line shows the ACF for LBGs from the whole box, corresponding errors on the ACF for LBGs are comparable with the line width, and black points represent the observational results by \citet{kashikawa2006}. Columns show the results for $f_{esc}=0.05,0.25$ and 0.5, as marked. The values of \avchi~ are marked at the end of each row, with the $f_{\alpha}/f_{c}$ value marked in each panel, along with the $\chi^2$ error.
  \label{fig_LAEclustering}}
\end{figure*}

The hydrodynamical simulation analysed in this work was carried out with the TreePM-SPH code GADGET-2 and has a box size of $80 h^{-1}$~comoving Mpc (cMpc) and contains $1024^3$ DM particles, and initially the same number of gas particles; the mass of a DM and gas particle is $3.6 \times 10^7 h^{-1} \Msun$ and $6.3 \times 10^6 h^{-1} \Msun$, respectively. The simulation includes all the standard processes of star formation and its associated metal production and feedback using the prescription of \citet{springel-hernquist2003b}, assuming a Salpeter \citep{salpeter1955} initial mass function (IMF) between $0.1-100 \Msun$. Bound structures of more than 20 total (Dark Matter, gas and star) particles are recognised as galaxies using the Amiga Halo Finder \citep[AHF;][]{knollmann2009}. Of all these galaxies, we only use ``resolved" galaxies that are complete in the halo mass function in all our calculations - these consist of at least 160 (10) gas (star) particles and have a halo mass $M_h \gsim 10^{9.2}\Msun$. Assuming each star particle to have formed in a burst, we calculate its spectra, and rest-frame intrinsic \Lya ($L_\alpha^{int}$) and UV continuum ($L_{\lambda,c}^{int}$; 1505\AA) luminosities depending on the stellar mass, age and metallicity using the population synthesis code STARBURST99 \citep{leitherer1999}. For each galaxy, the dust mass and its corresponding UV attenuation are calculated using the dust model described in \citet{dayal2011}. The observed UV specific luminosity is then calculated as $L_{\lambda,c}^{obs}=f_c\times L_{\lambda,c}^{int}$ where $f_c$ is the fraction of UV photons that emerge out of the ISM unattenuated by dust which is fixed by matching to the observed evolving UV LF at $z \simeq 6-8$; galaxies with an absolute UV magnitude $M_{UV} \lsim -17$ are then identified as LBGs. The observed \Lya luminosity ($L_\alpha^{obs}$) is then calculated as $L_{\alpha}^{obs} = f_{\alpha} T_{\alpha} L_{\alpha}^{int}$ where $f_\alpha$ and $T_\alpha$ account for dust attenuation in the ISM and by \HI in the IGM, respectively. Galaxies with $L_\alpha^{obs} \geq 10^{42}$~erg~s$^{-1}$ and $EW = L_\alpha^{obs}/L_{\lambda,c}^{obs} \geq 20$\,\AA\, are  identified as LAEs.

Given that both $L_\alpha^{int}$ and \avchi~depend on the fraction of \HI ionizing photons that can escape out of the ISM (\fesc), we use five values of \fesc=$0.05,0.25,0.5,0.75,0.95$ to post-process the $z\simeq 6.6$ snapshot of the hydrodynamical simulation with the 3D radiative transfer (pCRASH). To explore the full range of \avchi~(that determines $T_\alpha$) we run pCRASH starting from a completely neutral to a completely ionized state and obtain ionization fields for different values of the mean fraction of neutral hydrogen \avchi; for each galaxy we assume the initial line profile to be Gaussian and compute the average \Lya transmissions along 48 different lines of sight (LOS). For each combination of $f_{esc}$ and $\langle \chi_{HI} \rangle$ the transmission $T_{\alpha}$ is then fixed and the only free parameter that can be adjusted to match \Lya luminosities to observations is the relative escape fraction of \Lya and UV photons from the ISM, $p=f_{\alpha}/f_c$.  

From the \Lya LFs we find the following trends: the amplitude of the LAE \Lya LF decreases with increasing $f_{esc}$ due to a decrease in $L_\alpha^{int}$. However, this decline can be compensated by an increase in either $f_{\alpha}/f_c$ or $T_\alpha$ (in an increasingly ionized IGM). Comparing our model \Lya LFs to observations \citep{kashikawa2011} we found that allowing for clumped dust ($p\gsim 0.7$), the observations can be reproduced for \avchi$\simeq0.5-10^{-4}$, $f_{esc}\simeq0.05-0.5$  and $f_{\alpha}/f_c=0.6-1.8$ within a $1\sigma$ error.

\section{Constraints from LAE clustering}
\label{sec_LAE_clustering}
\begin{table}
\begin{center}
\caption{For the $\langle \chi_{HI} \rangle$ value shown in Column 1, we summarise the $f_\alpha/f_c$ ratio required to best fit the simulated \Lya LF to observations \citep{kashikawa2011} within $1\sigma$ limits in columns 2-4 with subscripts showing the \fesc value used \citep{hutter2014}.}
\begin{tabular}{|c|c|c|c|}
\hline
\avchi&$\langle f_{\alpha}$/$f_c\rangle_{0.05}$ &$\langle f_{\alpha}$/$f_c\rangle_{0.25}$ &$\langle f_{\alpha}$/$f_c\rangle_{0.5}$\\
\hline
0.50&1.2& 1.6& 1.8 \\
0.25&0.8&1.2 & 1.4\\
0.10&0.68 &1.0 & 1.4\\
0.01&0.68& 0.9 & 1.2\\
10$^{-4}$&0.60&0.8 & 1.2 \\
\hline
\end{tabular}
\label{table_bestfit}
\end{center}
\end{table}

\begin{figure*}
  \center{\includegraphics[width=1.03\textwidth]{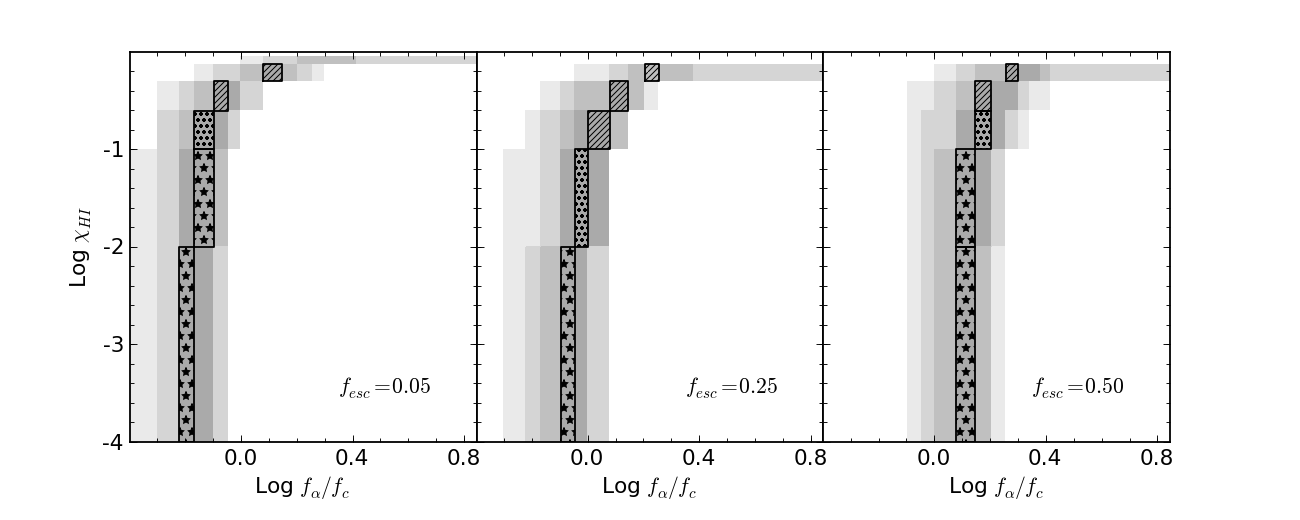}}
  \caption{The grey shaded regions show the $(1-5)\sigma$ (dark grey to white respectively) regions for the combinations of $f_{esc}$, $\langle \chi_{HI} \rangle$ and $f_{\alpha}/f_c$ that best fit the observed \Lya LF data from \citet{hutter2014}. We over-plot black contours for $3\sigma$ (stars), $5\sigma$ (dots), $>5\sigma$ (hatching) by comparing model results to observed ACF data \citet{kashikawa2006}. As shown, LAE clustering observations ($3\sigma$) require an IGM that has $\chi_{HI} \lsim 0.01$, $f_\alpha/f_c \leq 1.2$ and $f_{esc} \leq 0.5$. See Sec. \ref{sec_LAE_clustering} for details. \label{fig_LAE_chi}}
\end{figure*}

Given the sensitivity of \Lya photons to \HI, the ionization field should be imprinted in the spatial distribution of the LAEs which can be quantified by the two point correlation functions (ACF). In this section, we use the ACF of theoretical LAEs for each of these best-fit combinations (that match the \Lya LF to within a 1-$\sigma$ error as summarised in Table \ref{table_bestfit}) to narrow the joint constraints on \avchi, \fesc and $f_\alpha/f_c$. We start by describing our procedure for obtaining LAE ACFs, which depends both on the depth along the LOS \citep{peebles1980} since all galaxies are projected to a plane perpendicular to the LOS, as well as the chosen field of view (FoV), which should be comparable to that observed \citep{kashikawa2006}. Indeed, while the measured ACF should be independent of the FoV if large enough areas are sampled, the restricted observational FoV leads to an ACF that is not independent of sample variance. To get an estimate of the average ACF and its variance, we generate 36 mock catalogues (12 along each of $x,y,z$ directions) for over-lapping volumes comparable to that observed by \citet{kashikawa2006}, corresponding to a redshift distance $\Delta z\sim0.1$ at $z\simeq6.5$ and a field of view (FoV) of $\sim3\times10^3 h^{-2}$Mpc$^2$. Using the Landy-Szalay estimator we compute the ACF in each mock survey region denoted by $w_i(r)$ with respect to the mean LAE number density $\overline{n}$ of the complete simulation box and estimate the mean value of $w(r)$ as well as its variance from our mock catalogues as
\begin{eqnarray}
 \overline{n} (1+w(r))&=&\frac{1}{N} \sum_{i=1}^N n_i (1+w_i(r))
\end{eqnarray}

We start by calculating the LBG ACF (over the entire box) to get an estimate of the underlying galaxy population. As seen from Fig. \ref{fig_LAEclustering}, LBGs are almost homogeneously distributed and the ACF is consistent with essentially no clustering on scales $\lsim 30 h^{-1}$~Mpc. On the other hand, the LAE ACF is affected both by ISM dust, as well as the large scale topology of reionization. It might be expected that in the early stages of reionization, only those galaxies that are clustered and hence capable of building large \HII regions would be visible as LAEs (leading to a large amplitude of the ACF), with the amplitude of the ACF decreasing as reionization progresses and faint objects are able to transmit enough flux through the IGM to be visible as LAEs. Indeed, as shown in Fig. \ref{fig_LAEclustering}, LAEs exhibit precisely this behaviour. For a given $f_{esc}$ value (we remind the reader this corresponds to a fixed $L_\alpha^{int}$), as the IGM becomes more ionized (going down the vertical columns in the Fig. \ref{fig_LAEclustering}), smaller galaxies are able to transmit more of their flux through the IGM, requiring lowering $f_\alpha/f_c$ values to fit the \Lya LF. While for a half neutral IGM (panel a), only strongly clustered galaxies are visible as LAEs ($w(r) \simeq 4.5$) at scales of $\simeq 2 h^{-1}$Mpc, $T_\alpha$ increases for a completely ionized IGM at the same scale, leading to a more homogeneous LAE distribution resulting in a lower amplitude of the ACF (panel m; $w(r) \simeq 1.5$). 
 
At a given value of \avchi, $L_\alpha^{int}$ decreases with increasing $f_{esc}$ (horizontal rows in the same figure) which must be compensated by an increase in $f_\alpha/f_c$. However, this compensation results in very similar ACFs at a given \avchi~ value. Our results therefore show that the ACF is driven by the reionization topology (as determined by \avchi), with $f_{esc}$ and the local $f_\alpha/f_c$ having a marginal effect (a factor of about 1.5) on its amplitude. 

We then calculate the $\chi^2$ errors between our simulated ACFs and observations, and find that observations constrain \avchi$\lsim 0.01$ for $f_{esc} = 0.05,\ 0.5$ and \avchi$\lsim 10^{-4}$ for $f_{esc} = 0.25$ (to within a $3\sigma$ error). Further, while the $f_\alpha/f_c$ ratio is compatible with homogeneously distributed dust for $f_{esc}=0.05$, the decrease in $L_\alpha^{int}$ requires clumped dust ($f_\alpha/f_c \geq 0.7$) for $f_{esc}=0.25$ and 0.5.

To highlight the importance of the spatial clustering of LAEs, we show the $(1-5)\sigma$ constraints allowed by matching the \Lya LFs to observations in Fig. \ref{fig_LAE_chi}. As seen, these encompass a region such that \avchi$\simeq 10^{-4}-0.5$, $f_{esc}=0.05-0.5$ and $f_\alpha/f_c=0.6-1.8$. However, building ACFs for each of these allowed combinations, we find that theory and observations yield much tighter constraints of \avchi$\simeq 0.01-10^{-4}$, $f_{esc}=0.05-0.5$ and $f_\alpha/f_c=0.6-1.2$ to within a $3\sigma$ error. 

Finally, our results show that it is the reionization topology (as parameterised by \avchi) that drives the ACF, supporting the results obtained by \citet{mcquinn2007}, \citet{jensen2013} and \citet{jensen2014}. Although \citet{jensen2014} have assumed a simple scaling down of the \lya luminosity of all galaxies by a fixed amount due to galaxy evolution, they also find that galaxies are more likely to be observed as LAEs if they reside in ionized regions for \avchi$\gsim 20\%$. We note that the number of galaxies we identify as LAEs is similar to within 10\% for the different $f_{esc}$, \avchi and $f_\alpha/f_c$ combinations, demonstrating that the enhanced LAE clustering in our model can be attributed to an increasing neutral IGM.

We note that the average number of LAEs in our mock catalogues ($\sim300$) exceeds the number of identified objects in \citet{kashikawa2006}. In order to compare to a complete sample we have considered all galaxies with $L_{\alpha}^{obs}\ge10^{42}$erg~s$^{-1}$ and $EW\ge20$\AA~ as LAEs. Since this luminosity cut may include fainter galaxies than observed, the obtained ACFs represent a lower limit, as the clustering increases for higher luminosity cuts \citep{jensen2014}. 
According to \citet{jensen2013} our galaxy sample size of $\sim300$ is sufficient to distinguish a half-ionized from an ionized IGM, but does not provide the necessary sample size of $\sim500$ to distinguish \avchi$=0.25$ from \avchi$=1$. However, we have modelled the luminosity of each galaxy according to its stellar population obtained from the hydrodynamical simulation. This makes our results more sensitive to the ionization state of the IGM than \citet{jensen2013}; their clustering signal was reduced by the random scatter they added to the imposed fixed mass-to-light ratio. Hence, in agreement with \citet{mcquinn2007} we find our sample size sufficient to distinguish \avchi$=0.25$ from \avchi$=10^{-4}$.

\begin{figure*}
  \center{\includegraphics[width=1.0\textwidth]{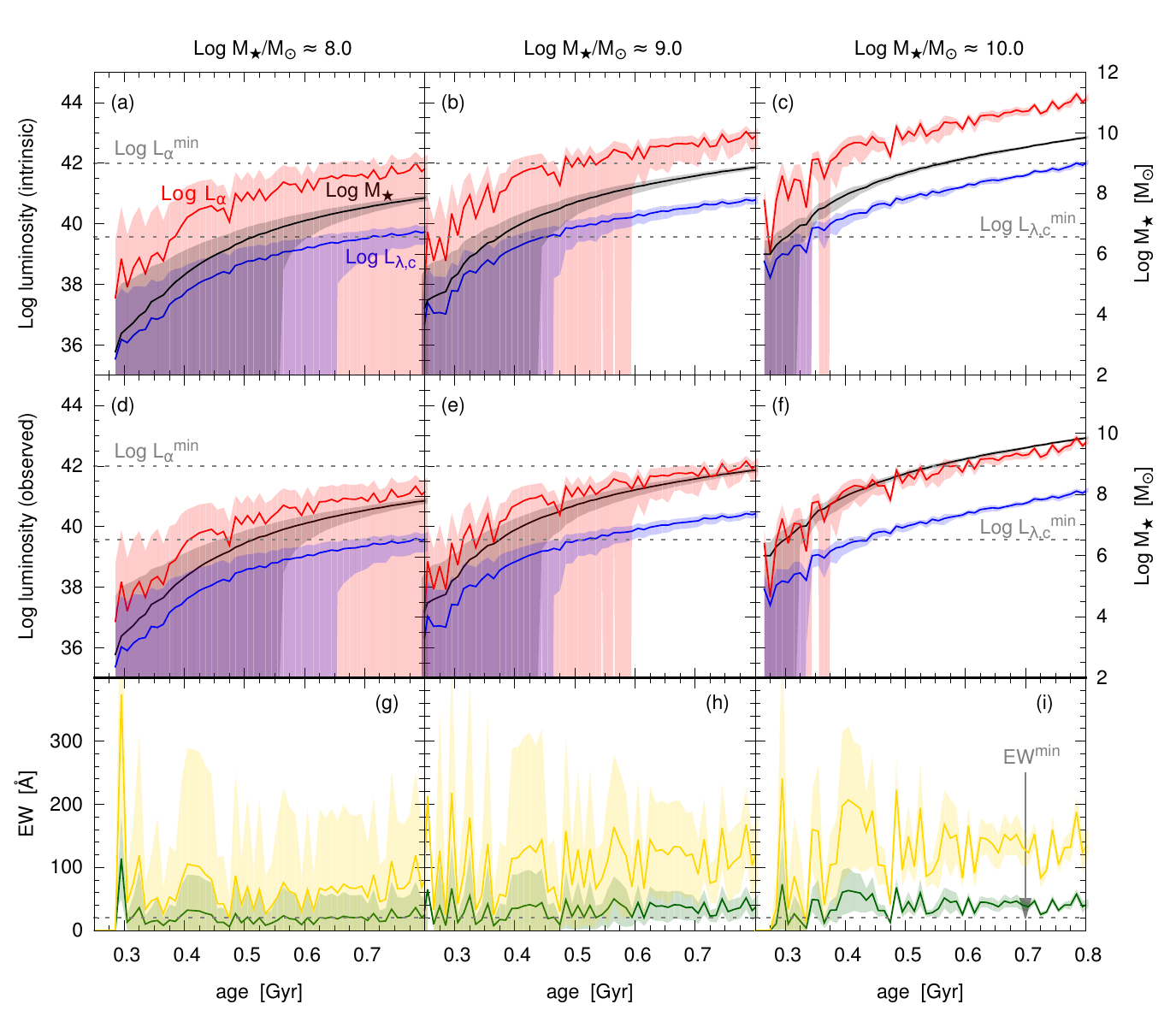}}
  \caption{Time evolution of $M_*$, \Lya and UV luminosities and EWs across three $M_*$ bins of $10^8, 10^9$ and $10^{10}\Msun$ as marked above each column. The upper 3 panels show the intrinsic quantities ($M_*$ in black; $L_\alpha^{int}$ in red (upper line); $L_{\lambda,c}^{int}$ in blue (lower line)); we use $f_{esc}=0$ for the \Lya luminosity. The middle three panels show the observed quantities ($M_*$ in black; $L_\alpha^{obs}$ in red (upper line); $L_{\lambda,c}^{obs}$ in blue (lower line)) where we assume $T_{\alpha}=0.45$, $f_{\alpha} = 0.68 f_c$ and a individual $f_c$ depending on the dust mass of each galaxy. The dashed lines in the top two panels show the current observational limits corresponding to the \Lya ($10^{42}$erg s$^{-1}$) and UV ($10^{39.6}$erg s$^{-1}/$\AA) luminosities. The lower 3 panels show the intrinsic (yellow, upper line) and observed (green, lower line) \Lya EWs with the dashed line showing the minimum limit of 20\AA. In each panel shaded regions show the variance in the given $M_*$ bin.
  \label{fig_hist_sources}}
\end{figure*}

\section{The relation between LAEs and LBGs}
\label{sec_LAE_LBG}
In this section we show how simulated $z \simeq 6.6$ galaxies build up their stellar mass ($M_*$), and the time evolution of their UV and \Lya luminosities, and \Lya equivalent widths (EW). We then present the fraction of time during the last $100$~Myrs prior to $z=6.6$ for which galaxies of different UV magnitudes are visible  as LBGs or as LBGs with detectable \Lya emission, both considering intrinsic and observed luminosities for each combination of $f_{esc}$, \avchi~ and $f_{\alpha}/f_c$ that reproduces the \Lya LF and LAE ACF within $1\sigma$ and $3\sigma$ respectively.
In order to consider a time span independent of resolution effects we have chosen a period of time of $100$~Myrs as an interval.
We discuss the limiting case, i.e. considering the total lifetime of each galaxy in our simulation as the corresponding period of time, in the Appendix \ref{a1}.

\subsection{Time evolution of stellar mass, Ly$\alpha$ and UV luminosities}
\label{subsec_Lya_UV}

\begin{figure*}
  \center{\includegraphics[width=0.4\textwidth,angle=270]{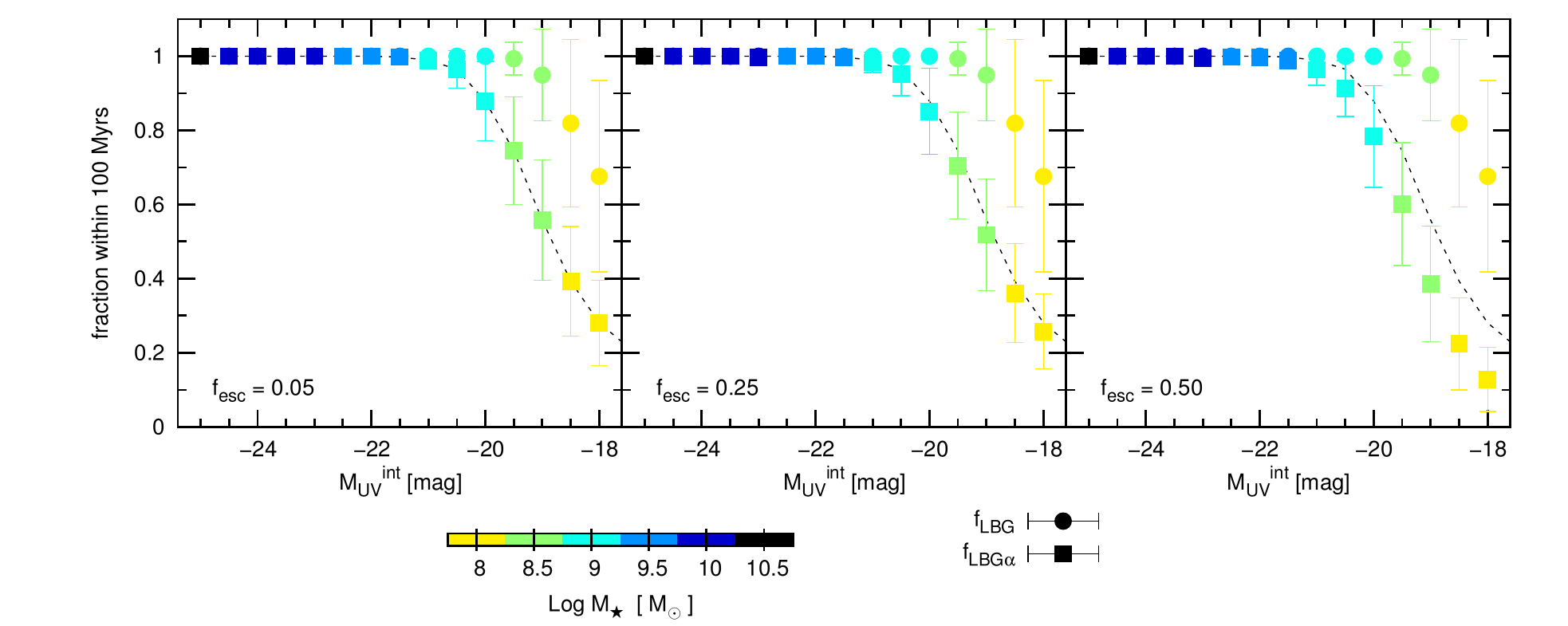}}
  \caption{ Fraction of time during the last $100$~Myrs prior to $z=6.6$ that galaxies spend as  LBGs ($f_{LBG}$, circles) and as LBGs with a \Lya line ($f_{LBG\alpha}$, squares), as a function of the intrinsic UV luminosity for intrinsic values of \Lya and UV luminosities. The panels show the fractions for the indicated values of $f_{esc}=0.05$, $0.25$, $0.5$. The mean stellar mass in each $M_{UV}$ bin is encoded in the shown colour scale. The fractions are computed as the mean of the galaxies within $M_{UV}$ bins $k$ ranging from $k-0.25$ to $k+0.25$ for $k=-25$~..~$-18$ in steps of $0.5$. Error bars show the standard deviations of the mean values. The dotted black line in each panel represents $f_{LBG\alpha}$ for $f_{esc}=0.05$; as clearly seen, increasing $f_{esc}$ to 0.5, leads to a decrease in $f_{LBG\alpha}$. \label{fig_bestfit_fractions_MUV_int}}
\end{figure*}

We use the ages of each star particle to trace the growth of $M_*$, and the metallicity and time dependent values of the intrinsic UV and \Lya luminosities ($L_{\lambda,c}^{int}$ and $L_\alpha^{int}$, respectively) and the intrinsic \Lya EW ($= L_{\alpha}^{int}/L_{\lambda,c}^{int}$) for galaxies in three bins of $M_* \simeq 10^{8,9,10}\Msun$. 
We assume that all ionizing photons emitted within galaxies are absorbed in the ISM and produce \Lya radiation, i.e. $f_{esc}=0$, to investigate the most extreme case.

As seen from the upper three panels of Fig. \ref{fig_hist_sources}, both $L_{\lambda,c}^{int}$ and $L_\alpha^{int}$ (averaged over the galaxies in the given $M_*$ bin) rise as galaxies steadily build up in (stellar) mass, albeit with a large scatter reflecting the assembly history of different galaxies. As expected for normal star forming galaxies, the average intrinsic \Lya EW for all the three $M_*$ bins considered here has a value between $30-300$\AA\, (lower 3 panels of the same figure) that is larger than the minimum value of 20\AA\, used to identify LAEs. Indeed, we find that once a galaxy exceeds a critical mass of roughly $10^{8.5} (10^{7.5}) \Msun$, it can produce enough luminosity to intrinsically be a LAE (LBG), it has also met the LBG criterion.

As for observed luminosities, we remind the reader that we compute $L_\alpha^{obs}$ using the fraction of UV photons that escape out of the galaxy ($f_c$); $L_\alpha^{obs}$ is computed assuming homogeneously distributed dust ($f_\alpha/f_c=0.68$) and an IGM transmission value of $T_\alpha = 0.45$ \citep{hutter2014}. As expected, including the effects of dust and the IGM reduces both the \Lya and UV luminosities, as shown in the middle panels of the same figure so that the critical $M_*$ at which a galaxy has the minimum luminosity to be a LAE (LBG) increases to  $10^{9.5} (10^{8.5}) \Msun$, although in most cases the observed EW is still larger than the minimum required value of 20\AA\, (bottom most panels). 

As expected from our discussion above, the time a galaxy spends as a LBG or LAE increases with increasing $M_*$. However, in our model a galaxy becomes a LBG before it also turns into a LAE because of the more stringent (luminosity + EW) constraints imposed on identifying a galaxy as a LAE, as expected from the critical $M_*$ values quoted above. 
To quantify, while galaxies with stellar masses of $\sim10^8\msun$ are visible as LBGs for roughly the last $80$~Myrs, they do not meet the selection criterion to be visible as LAEs: although their intrinsic EW are larger than the minimum value of 20\,\AA\ required by observations (bottom panel of same figure), $L_\alpha^{int}$ does not meet the required \Lya luminosity of $10^{42}$~erg~s$^{-1}$. 
Galaxies with $M_* \gsim 10^9 \Msun$ are massive enough to sustain star formation and maintain $L_{\lambda,c}^{int}$ and $L_\alpha^{int}$ values above the required limits. Again, the lifetime spent as a LAE is driven by $L_\alpha^{int}$, with the intrinsic EW always exceeding 20\,\AA (bottom panel of Fig. \ref{fig_hist_sources}). 

When considering observed luminosities, the time a galaxy is visible as a LBG decreases; the decrease is more pronounced for LAEs that are additionally affected by IGM transmission. From panels (d)-(f) of Fig. \ref{fig_hist_sources}, we see that in addition to galaxies with $10^8 \Msun$, galaxies with $M_* \simeq 10^9 \Msun$ are also no longer visible as LAEs as a result of the \Lya luminosity dropping below visible limits.

\subsection{The fraction of time spent as LBG and as LBG with \Lya emission}
\label{subsec_duty_cycle}

\begin{figure}
  \center{\includegraphics[width=0.6\textwidth,angle=270]{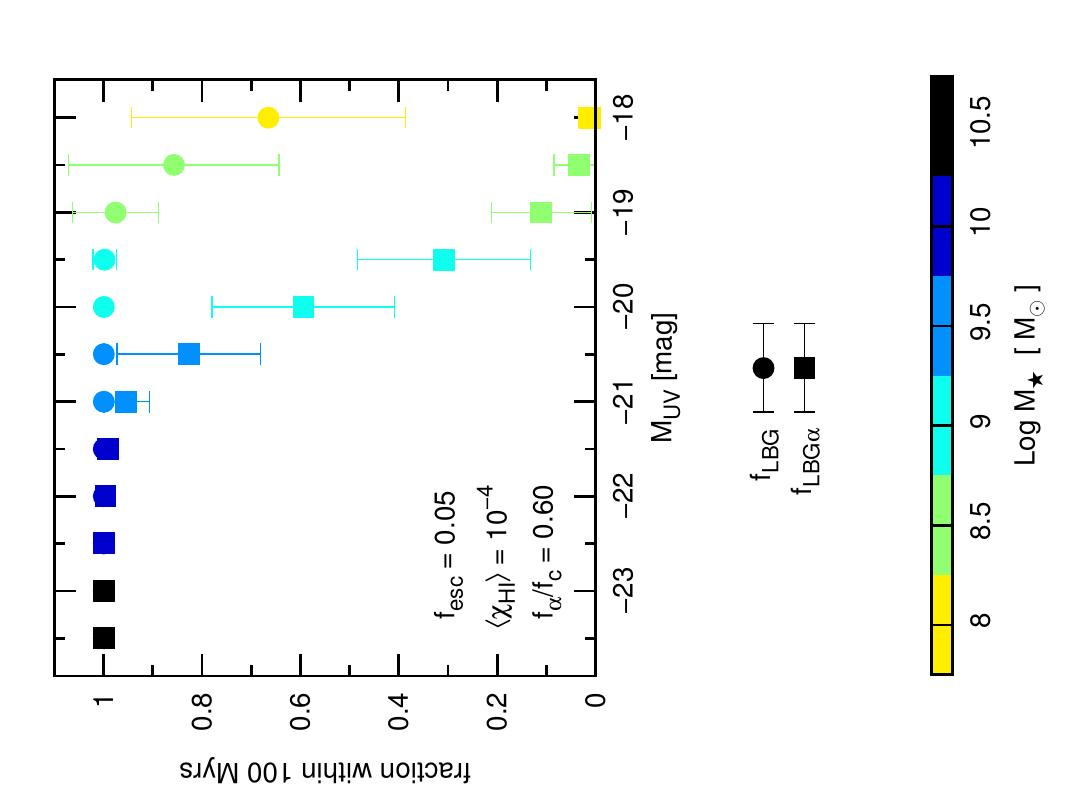}}
  \caption{ Fraction of time during the last $100$~Myrs prior to $z=6.6$ that galaxies spend as LBGs ($f_{LBG}$, circles) and as LBGs with a \Lya line ($f_{LBG\alpha}$, squares) as a function of the UV luminosity for our best fit models. The mean stellar mass in each $M_{UV}$ bin is encoded in the shown colour scale. This panel shows the best fit case for $f_{esc}=0.05$, \avchi$=10^{-4}$ and $f_{\alpha}/f_c=0.60$. The fractions are computed as the mean of the galaxies within $M_{UV}$ bins $k$ ranging from $k-0.25$ to $k+0.25$ for $k=-25$~..~$-18$ in steps of $0.5$. Error bars show the standard deviations of the mean values. The independence of $f_{LBG\alpha}$ on the chosen best fit model clearly shows that the effects of $f_{esc}$, \avchi~ and $f_{\alpha}/f_c$ are degenerate on $f_{LBG\alpha}$.\label{fig_bestfit_fractions_MUV}}
\end{figure} 
We now calculate the fraction of time during the last $100$~Myrs prior to $z=6.6$ that galaxies in different UV magnitude bins spend as a LBG ($f_{LBG}$) and as a LBG with \Lya emission ($f_{LBG\alpha}$, i.e. as a LAE), and discuss the resulting fractions ($f_{LBG\alpha}$ and $f_{LBG}$) for galaxies of varying UV magnitudes $M_{UV} = -18$ to $-25$ in more detail.
We start with intrinsic UV luminosities as shown in Fig. \ref{fig_bestfit_fractions_MUV_int}: firstly, as galaxies become more massive, they are able to sustain large rates of star formation, leading to $M_{UV}$ values that scale with $M_*$. Secondly, we find that both $f_{LBG}$ and $f_{LBG\alpha}$ increase with an increase in the UV magnitude (or $M_*$), as explained in Sec. \ref{subsec_Lya_UV} above: $f_{LBG}$ increases from 65\% to 100\% as $M_*$ increases from $10^{8}-10^{10.5}\Msun$. As a result of the stricter luminosity and EW criterion imposed to identify galaxies as LAEs, $f_{LBG\alpha}<f_{LBG}$ and declines more rapidly than $f_{LBG}$ towards fainter UV luminosities, decreasing from 100\% to 10-30\% as $M_*$ decreases from $10^{10.5}-10^8 \Msun$. Further, the decrease in $L_\alpha^{int}$ with increasing \fesc results in a (linear) decrease in $f_{LAE}$ as shown from panels (a)-(c) of the same figure: while $f_{LAE}$ is comparable for $f_{esc}=0.05,0.25$, it decreases by about $0.15$ for $f_{esc}=0.5$, where half of the ionizing photons do not contribute to the \Lya luminosity thereby reducing the fraction of time it shows \Lya emission.

We then calculate $f_{LBG}$ and $f_{LBG\alpha}$ including the effects of dust and IGM attenuation for each of the best-fit parameter combinations that match both the observed \Lya LF and ACF as shown in Sec. \ref{sec_LAE_clustering}. We use the $f_c$ value for each galaxy according to its final dust mass at $z\simeq6.6$, the \Lya transmission $T_{\alpha}$ of each galaxy was obtained from the ionization field and the ratio of the escape fractions of \Lya and UV continuum photons ($f_{\alpha}/f_c$) was set according to Table \ref{table_bestfit}. As can be seen from Fig. \ref{fig_bestfit_fractions_MUV} (since $f_{LBG\alpha}$ is nearly identical for all best fit cases, we show only one best fit case) the additional attenuation by dust in the ISM and neutral hydrogen in the IGM leads to a rise of the mean stellar mass in each $M_{UV}$ bin, as well as to lower values for $f_{LBG\alpha}$ (compared to the intrinsic case) but not for $f_{LBG}$. Nevertheless, we find the same trends as when considering intrinsic luminosities: $f_{LBG}$ always exceeds $f_{LBG\alpha}$, and $f_{LBG}$ and $f_{LBG\alpha}$ decrease towards fainter UV luminosities with $f_{LBG\alpha}$ declining more rapidly. However the relative difference between $f_{LBG}$ and $f_{LBG\alpha}$ is larger and the decline of $f_{LBG\alpha}$ is more rapid: while $f_{LBG}$ decreases from 100\% to 65\% as $M_{UV}$ increases from -23.5 to -18.5, $f_{LBG\alpha}$ drops from 100\% to essentially 0 for the same magnitude range. The stronger decline in $f_{LBG\alpha}$ is not only due to the additional dust- and IGM attenuation of the \Lya luminosities, but also due to the additional selection criterion in \Lya equivalent width which becomes more important towards UV fainter galaxies. 

We also find that the decrease in $f_{esc}$ is compensated by the increase in $f_\alpha/f_c$, leading to very similar $f_{LBG,\alpha}$ ratios for all the models
; note that the IGM is almost ionized in most cases, leading to similar $T_\alpha$ values. This clearly shows that \avchi, $f_{esc}$ and $f_\alpha/f_c$ compensate each other \citep[as shown in][]{hutter2014}, as a result of which $f_{LBG\alpha}$ only depends on the combination of the parameters but not on their individual values.

Thus in our model, the most luminous (massive) LBGs most often show \Lya emission, irrespective of whether intrinsic or observed luminosities are considered.

\section{Discussion \& Conclusions}
\label{sec_conclusions}
\begin{figure}
  \center{\includegraphics[width=0.5\textwidth]{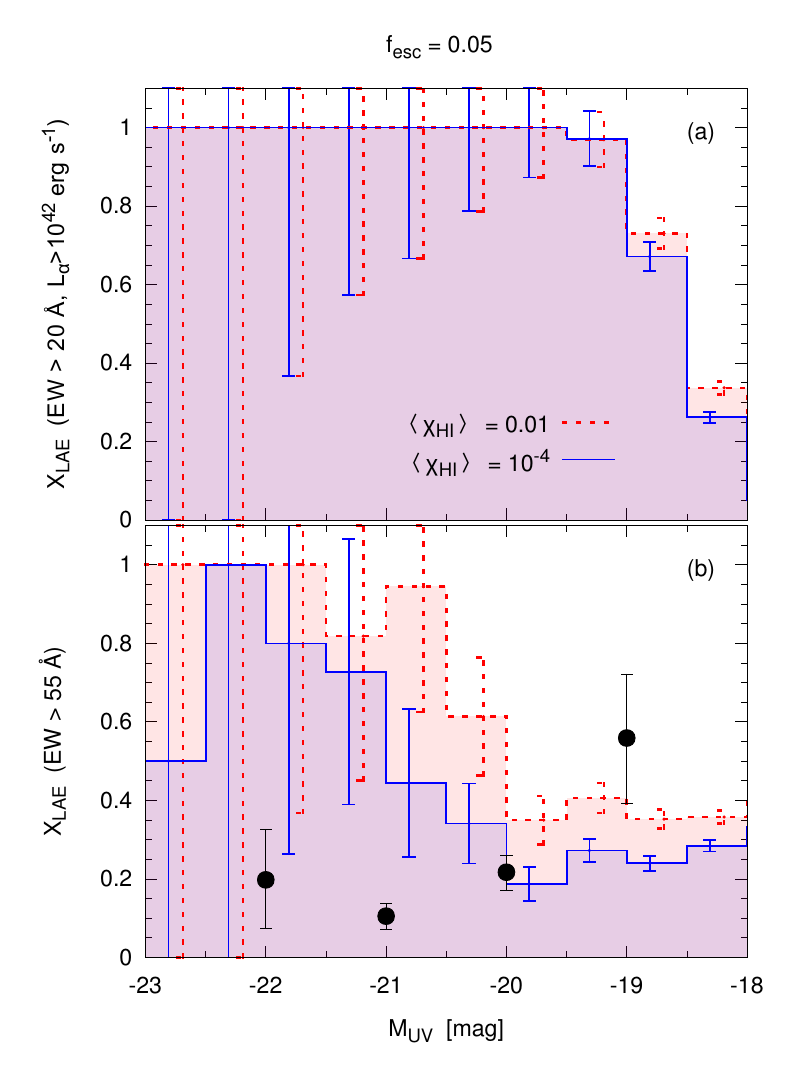}}
  \caption{ Fraction of LBGs detected also as LAE ($X_{LAE}$) as a function of UV luminosity. The fractions are shown for fiducial best fit models of $f_{esc}=0.05$: \avchi$=0.01$, $f_{\alpha}/f_c=0.68$ (red dashed) and \avchi$=10^{-4}$, $f_{\alpha}/f_c=0.60$ (blue solid). Error bars are derived from the Poissonian errors of the LAE and LBG numbers. The upper panel (a) shows the fraction $X_{LAE}$ for a LAE selection criterion of $L_{\alpha}>10^{42}$erg~s$^{-1}$ and $EW>20$\AA, and the lower panel (b) adopts an selection criterion of $EW>55$\AA, which is in agreement with the observations by \citet{stark2010}. Black points represent the observational constraints of \citet{stark2010} at $4.5<z<6.0$.\label{fig_MUV_fracLAE}}
\end{figure} 

We couple a cosmological hydrodynamical simulation (GADGET-2) with a dust model and a radiative transfer code (pCRASH) to model high-$z$ LAEs. Starting from a neutral IGM, we run pCRASH until the IGM is completely ionized, for $f_{esc}$ values ranging from $0.05$ to $0.95$. In \citet{hutter2014}, we showed that comparing model results to \Lya LF observations simultaneously constrains the escape fraction of ionizing photons $f_{esc}$, the mean amount of neutral hydrogen \avchi~ and the ratio of the escape fractions of \Lya photons and UV continuum photons $f_{\alpha}/f_c$ to \avchi$\simeq 10^{-4}-0.5$, $f_{esc}=0.05-0.5$ and $f_\alpha/f_c=0.6-1.8$. In this paper, we calculate the ACFs for these different combinations and find that comparing these to observations significantly narrows the allowed 3D parameter space (within a $3\sigma$ error) to \avchi$\simeq 0.01-10^{-4}$, $f_{esc}=0.05-0.5$ and $f_{\alpha}/f_c=0.6-1.2$. While the effects of these three parameters are degenerate on the \Lya LFs, the ACF is most sensitive to large-scale ionization topologies and reionization leaves clearly distinguishable ACF imprints (boosting up the strength of the ACF) that can not be compensated by varying $f_{esc}$ or $f_\alpha/f_c$. Further, the ACF allows us to constrain \avchi$\leq0.01$, independent of the other two parameters, and we also constrain $f_{esc}\leq0.5$ and $f_{\alpha}/f_c\leq1.2$.

We then analyse the average time evolution of UV and \Lya luminosities of simulated galaxies at $z \simeq 6.6$ in three bins of $M_* = 10^{8,9,10}\Msun$, finding the following: as soon as a galaxy exceeds a critical stellar mass of $M_* \simeq 10^{8.5} (10^{7.5}) \Msun$, its intrinsic \Lya (UV) luminosity is large enough for it to be identified as a LAE (LBG). Including the effects of dust and IGM attenuation naturally results in an increase in this critical mass to $10^{9.5}$ and $10^{8.5} \Msun$ for LAEs and LBGs respectively. 

Considering the fraction of time during the last $100$~Myrs (prior to $z=6.6$) a galaxy spends  as a LBG with \Lya emission ($f_{LBG\alpha}$) or as a LBG ($f_{LBG}$), we find that the former is always smaller due to the more stringent luminosity and EW constraints imposed on identifying galaxies as LAEs. We find that both the intrinsic and dust-attenuated fraction $f_{LBG\alpha}$ and $f_{LBG}$ rise with increasing UV luminosity (and hence $M_*$): intrinsically, $f_{LBG}$ ($f_{LBG\alpha}$) increases from 65\% to 100\% (10-30\% to 100\%) as $M_*$ increases from $10^{8}-10^{10.5}\Msun$.  As expected, including the effects of dust and IGM transmission reduces the values for $f_{LBG\alpha}$ such that $f_{LBG\alpha}$ decreases from 100\% to essentially 0 as $M_*$ decreases from $10^{10.5} - 10^8\Msun$. Finally, we find that the fraction $f_{LBG\alpha}$ of all our models that reproduce the observed \Lya LF and LAE ACFs are independent of the chosen set of parameters: a larger $f_\alpha/f_c$ compensates a decrease in $T_\alpha$, or an increase in $f_{esc}$. As a result, $f_{LBG\alpha}$ only depends on the combination of these 3 parameters but not on their individual values. Thus, it is most often the most luminous LBGs that are visible in the \Lya.

Finally, we summarise the major caveats involved in this study. 
Firstly, given the cosmological volumes probed by the simulation, we are unable to resolve Lyman Limit systems (LLS) which could lead to a further decrease in the transmission $T_\alpha$ along lines of sight (LOS) intercepted by such systems \citep{bolton_haehnelt2013}. However, whether LLS are preferentially located in clustered regions, leading to an increasing suppression of $T_{\alpha}$ for massive galaxies remains an open question. 

Secondly, as a natural consequence of simulating cosmological volumes we are unable to resolve the ISM of individual galaxies, for which reason we assume a Gaussian profile (with a width set by the rotation velocity of the galaxy) for the Ly$\alpha$ line that emerges out of any galaxy which is probably an unrealistic scenario \citep[see e.g.][]{verhamme2008}. We note that our constraint on the ionization state of the IGM is model dependent, since the IGM \Lya transmission is sensitive to the the assumed line profile \citep{jensen2013}.

Thirdly, we assume dust attenuation and IGM transmission to be equal to the values at $z\simeq6.6$ in order to calculate both $f_{LBG\alpha}$ and $f_{LBG}$. While the dust mass (and hence attenuation) would be expected to be lower at earlier times, tracing this buildup would require tracking the dust growth in the progenitors of our simulated galaxies. This is beyond the scope of our present paper and we defer to this analysis to a work that is in preparation. Fourthly, $f_{LBG\alpha}$ would be expected to decrease with increasing $z$ as a result of an increase in \avchi~ (leading to a decrease in $T_\alpha$). However, properly accounting for the latter effect requires modelling the entire history of reionization.  

While we have explored the full range of possible values for $f_{esc}$, its mass and $z$-dependence remain poorly understood, which is also one of the main caveats involved in modelling the time-evolution of reionization. An increase in $f_{esc}$ with decreasing mass \citep[e.g.]{ferrara2013} would suppress the visibility of low-mass objects, and strongly impact the reionization fields we generate, emphasising the strong clustering of high-mass halos \citep{kaiser1984,bardeen1986,mo_white1996}, whilst depressing the LF at the faint end. Bringing these values into agreement with observations would then require an $f_{\alpha}/f_c$ ratio that decreases with increasing mass. With its observations of the ionization topology, instruments such as LoFAR will be invaluable in answering some of these outstanding questions.

Finally, we calculate the fraction of LBGs that would be identified as LAEs, $X_{LAE}$. Imposing a LAE selection criterion of $L_{\alpha}\ge10^{42}$erg~s$^{-1}$ and $EW\ge20$\AA~ we find that it is the faintest LBGs that do not show Ly$\alpha$ emission \citep[cf.][]{dayal2012}, while all LBGs brighter than $M_{UV}\sim-20$ are identified as LAEs at $z\simeq6.6$ (see Fig. \ref{fig_MUV_fracLAE}). Even if the Ly$\alpha$ selection criterion is made more stringent, i.e. $EW>55$\AA, $X_{LAE}$ does not show the behaviour observed by \citet{stark2010}, increasing instead of decreasing with UV magnitude. 
This mis-match is probably due to a combination of the following effects: 
firstly, the Ly$\alpha$ IGM transmission ($T_{\alpha}$) is subject to a large variance along different lines of sight (LOS) \citep[see][]{hutter2014} due to the patchy nature of reionization; while we use $T_\alpha$ values averaged over 48 LOS, using values along a specific LOS would lead to an over- or underestimate of $T_{\alpha}$. 
Secondly, as mentioned before, it is possible that the inclusion of LLS could decrease $T_{\alpha}$ of massive galaxies, leading to lower \Lya EW values respectively. 
Thirdly, the results by \citet{stark2010} are based on a post-reionization sample of galaxies ($4.5<z<6.0$), while our simulation samples the end of reionization era. The evolution of the UV (and \Lya) LFs of LAEs and LBGs suggest that the emitted observable \Lya radiation varies with cosmic epoch, depending predominantly on the evolution of dust and gas at $z<6$ and on the IGM neutral hydrogen fraction at $z>6$. Thus, the probe of LBGs with/without Ly$\alpha$ emission in \citet{stark2010} (dusty, ionized IGM) may differ to the LBG and LAE population at the end of reionization (less dusty, partly neutral IGM). Indeed, a remarkable high fraction of strong LAEs among $6.0<z<6.5$ luminous LBGs ($-21.75<M_{UV}<-20.25$) \citep{curtislake2012} indicates that most LBGs are also identified as LAEs, which is in agreement with our findings. However, their sample suffers from low statistics, which will be overcome by upcoming Subaru/HST/UltraVISTA data.
Finally, as an observational caveat, \citet{verhamme2012} have shown that detailed Ly$\alpha$ radiative transfer calculations of simulated galaxies suggest stronger inclination effects for Ly$\alpha$ photons than for UV continuum photons, introducing biases in the selection function of narrow-band LAE surveys that could lead to a significant fraction of LBGs galaxies being missed as LAEs. We aim at investigating these effects in detail in future works and shedding light on the tantalising connection between LBGs visible/invisible as LAEs.

\section*{Acknowledgements} 

The authors thank the anonymous referee for their insightful comments and suggestions that helped to improve the paper, as well as M. Dijkstra, N. Kashikawa, D. Schaerer and P. Creasey for useful discussions. 
PD acknowledges the support of the Addison Wheeler Fellowship awarded by the Institute of Advanced Study at Durham University.

 
\bibliographystyle{mn2e}
\bibliography{ew}

\appendix

\section{The fraction of lifetime spent as LBG with and without Ly$\alpha$ emission} 
\label{a1}
\begin{figure}
  \center{\includegraphics[width=0.6\textwidth,angle=270]{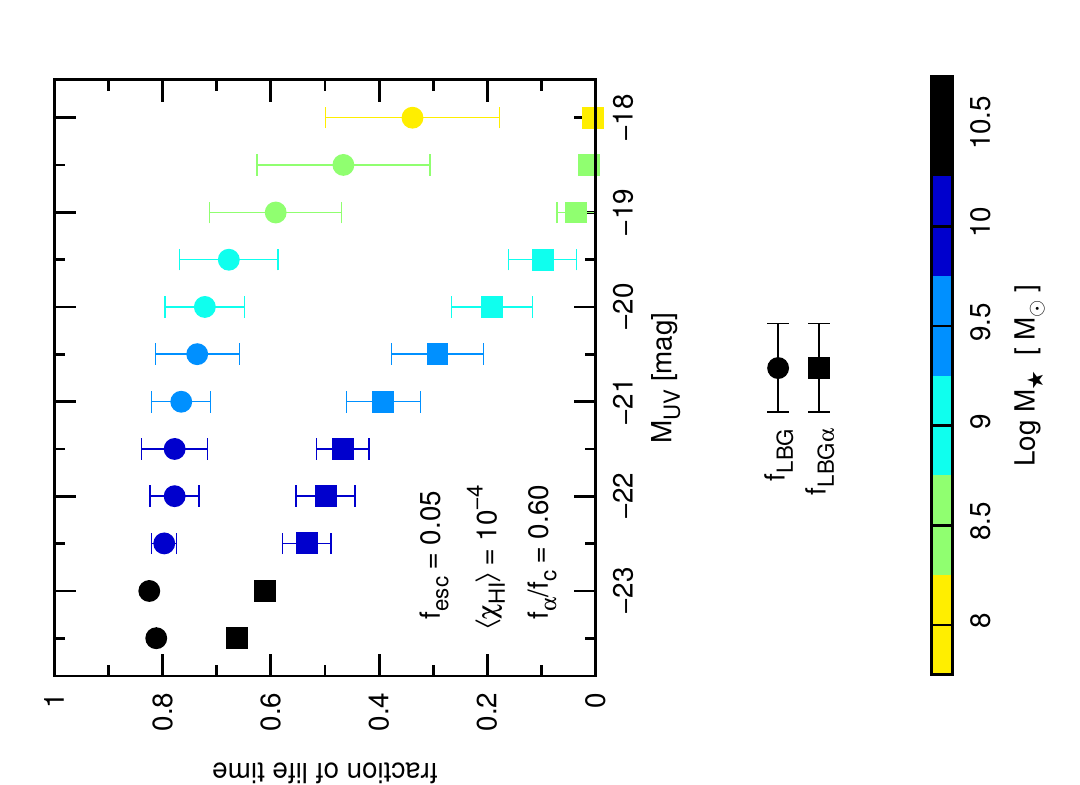}}
  \caption{Fraction of lifetime that galaxies spend as LBGs ($f_{LBG}$, circles) or as LBGs and LAEs ($f_{LBG\alpha}$, squares) as a function of the UV luminosity for our best fit models. The mean stellar mass in each $M_{UV}$ bin is encoded in the shown colour scale.  This panel shows the best fit case for $f_{esc}=0.05$, \avchi$=10^{-4}$ and $f_{\alpha}/f_c=0.60$. We omit to show the other best fit cases, since they are nearly identical to the shown one. For each best fit model we assume individual $f_c$ values for each galaxy according to its final dust mass at $z\simeq6.6$; the \Lya transmission $T_{\alpha}$ of each galaxy was obtained from the respective ionization field and the ratio of the escape fractions of \Lya and UV continuum photons was set according to Table \ref{table_bestfit}. The fractions are computed as the mean of the galaxies within $M_{UV}$ bins $k$ ranging from $k-0.25$ to $k+0.25$ for $k=-25$~..~$-18$ in steps of $0.5$. Error bars show the standard deviations of the mean values. \label{fig_bestfit_fractions_MUV_total}}
\end{figure} 

We also show results for the limiting case where the age is the time since the onset of star formation, showing how the fractions of lifetime change when the time span is increased. 
We take the total age of a galaxy as the time between the formation of the first star and $z=6.6$, and calculate the corresponding fractions of lifetime that galaxies in different $M_{UV}$ bins spend as a LBG and as a LBG with \Lya emission ($f_{LBG}$ and $f_{LBG\alpha}$, respectively).
As described in Section \ref{subsec_duty_cycle}, we compute the fractions of lifetime for our best fit models (see Table \ref{table_bestfit}) and show the fractions for the $f_{esc}=0.5$, \avchi$\simeq0.01$ and $f_{\alpha}/f_c$ best fit case in Fig. \ref{fig_bestfit_fractions_MUV_total}. We note that the fractions $f_{LBG\alpha}$ of all other best fit cases are nearly identical to the shown one.
Compared to the fractions derived for a time span of $100$~Myrs, the fractions of the total lifetime decrease for all UV luminosities. Galaxies that have been identified as LBGs or LBGs with Ly$\alpha$ emission within $100$~Myrs prior to $z=6.6$ have been fainter at earlier times where they have not met the selection criteria. The decrease in $f_{LBG}$ and $f_{LBG\alpha}$ illustrates the growth of stellar mass in galaxies with time and the existence of a critical stellar mass above which a galaxy produces enough luminosity to be identified as a LBG or as a LBG with Ly$\alpha$ emission. However, we like to note that the point in time when the first star forms depends on the resolution of the simulation, making our fractions of lifetime resolution-dependent but showing their trends as the time span is increased.

\end{document}